\documentclass[prd,twocolumn,superscriptaddress]{revtex4}

\newcommand{\bw}{\begin{widetext}}
\newcommand{\ew}{\end{widetext}}

\usepackage{graphicx}
\usepackage{amssymb,amsmath,amsbsy}
\usepackage{hyperref}
\newcommand{\be}{\begin{equation}}
\newcommand{\ee}{\end{equation}}
\newcommand{\bea}{\begin{eqnarray}}
\newcommand{\eea}{\end{eqnarray}}

\newcommand{\bracket}[2]{\bra{#1}\,#2\rangle} 
\newcommand{\bra}[1]{\langle\,#1\,|}          
\newcommand{\ket}[1]{|\,#1\,\rangle}          

\newcommand{\p}{\partial} 
\newcommand{\ud}{\mathrm{d}} 

\newcommand{\A}{\mathcal{A}}
\newcommand{\e}{\mathrm{e}} 
\newcommand{\x}{\boldsymbol{x}}
\newcommand{\y}{\boldsymbol{y}}
\newcommand{\z}{\boldsymbol{z}}
\newcommand{\n}{\boldsymbol{n}}
\newcommand{\pb}{\boldsymbol{p}}
\newcommand{\vc}[1]{\mbox{\boldmath$#1$}}

\newcommand{\ssvc}[1]{\mbox{\scriptsize\boldmath$#1$}}

\newcommand{\hbo}{\hbox to 1 true cm {\hfill } }
\newcommand{\tr}{\hbox{tr}}

\begin{document} 
\title{Probing the ground state in gauge theories}

\author{T.~Heinzl}\email[]{theinzl@plymouth.ac.uk}
\author{A.~Ilderton}\email[]{abilderton@plymouth.ac.uk}
\author{K.~Langfeld}\email[]{klangfeld@plymouth.ac.uk}
\author{M.~Lavelle}\email[]{mlavelle@plymouth.ac.uk}
\affiliation{School of Mathematics \& Statistics, University of Plymouth, Plymouth, PL4 8AA, UK}
\author{W.~Lutz}\email[]{wlutz@plymouth.ac.uk}
\affiliation{School of Mathematics \& Statistics, University of Plymouth, Plymouth, PL4 8AA, UK}
\affiliation{Institut f\"ur Theoretische Physik, Universit\"at T\"ubingen, Auf der Morgenstelle 14, D-72076 T\"ubingen, Germany} 
\author{D.~McMullan}\email[]{dmcmullan@plymouth.ac.uk}
\affiliation{School of Mathematics \& Statistics, University of Plymouth, Plymouth, PL4 8AA, UK}

\begin{abstract}
We consider two very different models of the flux tube
linking two heavy quarks: a string linking the matter fields and a
Coulombic description of two separately gauge invariant charges. We
compare how close they are to the unknown true ground state in compact U(1) and the SU(2) Higgs model. Simulations
in compact U(1) show that the string description is better in the
confined phase but the Coulombic description is best in the
deconfined phase; the last result is shown to agree with analytical
calculations. Surprisingly in the non-abelian theory the Coulombic
description is better in both the Higgs and confined phases. This
indicates a significant difference in the width of the flux tubes in
the two theories.
\end{abstract}

\maketitle

\section{Introduction}
We can naively think of a heavy meson as being made up of a static
quark and a static anti-quark separated by some distance. Around
these fermions there will be a cloud of glue whose form is unknown but is
typically thought of as being \lq cigar shaped\rq\ (we neglect
light quark flavours in this paper).

It is important to realise that the matter fields in an interacting
gauge theory, which are not locally gauge invariant, cannot be
directly interpreted as observables. Thus a crucial role of the glue
around quarks is to make the system gauge invariant. It is easy to
imagine doing this by linking the fermions by a string like gluonic
line. In an operator approach this is the familiar path ordered
exponential which, assuming it to have a non-zero overlap with the
ground state, is evolved in time to form the rectangular Wilson loop
approach to the interquark potential.

It is worth noting that in such a string like state the only gauge
invariant object is the overall colourless meson state. There are no
separately gauge invariant constituent quarks and, even at small
interquark separations, there is no sign of a Coulombic potential.

While producing a confining potential might be thought to compensate
for this deficit, it is well known that lattice simulations are
greatly improved by \lq smearing\rq\ the string. This
improvement is due to the naive Wilson loop being formed from an
initial state where the flux is trapped on an infinitesimally thin
path. As we will see below in U(1) theory, where we can calculate
analytically, this leads to the stringy description actually having
zero overlap with the true ground state unless a UV cutoff (such as
a finite lattice spacing) is imposed.

An alternative description, which may be expected to be closer to the
ground state at shorter separations in QCD, is to `dress' the quarks
separately so that they are each locally gauge invariant. For static
quarks this may be done via a Coulombic dressing which has been
shown \cite{Bagan:2005qg} to generate the perturbative interquark potential (parts
of which have been verified to next to next to leading order). It
also allows an identification of the gluonic structures which
generate the anti-screening and screening interactions, and incorporating this dressing removes
perturbative infra-red divergences in on-shell Green's functions.
Beyond perturbation theory the Gribov ambiguity sets a fundamental limit \cite{Ilderton:2007qy} on the observability
of individual dressed quarks, but it has previously \cite{Heinzl:2007cp} been shown that such Coulombic states may nevertheless be used on the lattice simply by
averaging over Gribov copies.

In this paper we will compare the  string like (or `axial') and
Coulombic descriptions of a heavy quark-antiquark  system. We will
do this analytically for deconfined U(1) and on a lattice for both compact U(1) and the SU(2) Higgs model. We will see that there are great differences in the two theories.

We make contact between our analytic and numerical studies of the ground state through a particular probe, namely the ratio
\be
    \frac{|\bracket{\psi'}{0}|^2}{|\bracket{\psi}{0}|^2}\; ,
\ee 
for any two states $\ket{\psi}$ and $\ket{\psi'}$ and where
$\ket{0}$ is the (typically unknown) ground state of the theory.
Simply, if this ratio is less than one then the overlap of the
ground state with $\ket{\psi}$ is better than the overlap with
$\ket{\psi'}$, and we should expect $\ket{\psi}$ to yield the
closer description of physics in the ground state. If the ratio is
greater than one then $\ket{\psi'}$ has the better overlap.

This ratio can be realised both analytically, for the deconfined U(1) theory, and on the
lattice through calculation of the large Euclidean time limit
\be
    \lim_{t\to\infty}\frac{\bra{\psi'}\e^{-\hat{H}t}
    \ket{\psi'}}{\bra{\psi}\e^{-\hat{H}t}\ket{\psi}}\; ,
\ee
since in Euclidean space the large time limit is a vacuum projector \cite{Heinzl:2007ca}.

This paper is organised as follows. We begin in Section \ref{analytic} by analysing a system of two non-confined abelian charges. This explicit investigation will allow us
to build up ideas and methods which will be applicable more
generally. We will see that the Coulombic
description yields the ground state while the axial system does not
yield the correct potential and is severely divergent. This is in
good agreement with physical expectations. In Section \ref{U1sect} we describe the lattice techniques we employ and go on to examine the ground state in compact U(1) in both the
deconfined and confining phases. It will be shown that the
simulations in the deconfined phase are in good agreement with the 
analytical results. In Section \ref{SU2sect} we apply our lattice techniques to an SU(2) Higgs theory. Finally
in Section \ref{concs} we give our conclusions.

\section{Analytic treatment of deconfined U(1) theory}\label{analytic}
In this section we consider QED with very heavy fermions, identified with static external sources. This is an ideal testing ground for our methods, as we may solve this theory exactly and analytically. This will allow us to build up an intuition for the physics involved before proceeding to more complex systems. The identification of the heavy fermions with physical objects is, however, somewhat suspect, as the Lagrangian fermions are not gauge invariant. Creating physical charges in gauge theories is the focus of the dressing approach, to be discussed below.
We work in Weyl gauge, $A^0=0$, \cite{Weyl:1950} and with very heavy fermions our Hamiltonian is simply that of free gauge fields,
\be \label{U1H}
  H = \frac{1}{2} \int d^3 x \, (\vc{E}^2 +
  \vc{B}^2) \; ,
\ee
where $\vc{B}$ is the magnetic field. Quantisation in Weyl gauge is straightforward \cite{Weyl:1950,Jackiw:1979ur,Jackiw:1983nv} as gauge potentials and electric fields are canonical variables. Hence we impose the non-zero
equal time commutator
\be\label{comm}
	  [\hat{A}_i(\x,t), \hat{E}_j(\y,t)] = i\delta_{ij} \delta^3 (\x - \y) \; .
\ee
We will work in the Schr\"odinger representation where the commutator (\ref{comm}) is realised on the state space by diagonalising the field operator $\hat{A}_i$ and taking $\hat{E}_i$ to act as a derivative, in direct analogy to the position representation of quantum mechanics. States $\ket{\Psi,t}$ are then identified with functionals $\Psi[\vc{\A},t]$ of a vector field $\vc{\A}$ on which the operators act as 
\be\begin{split}
  \bra{\vc{\A}}\hat{A}_i(\x)\ket{\Psi,t} &= \A_i(\x) \Psi[\vc{\A}, t]\;,\\
  \bra{\vc{\A}}\hat{E}_i(\x) \ket{\Psi,t} &= -i \frac{\delta}{\delta
  \A_i(\x)} \Psi[ \vc{\A}, t] \; .
\end{split}\ee
As we quantise the full gauge potential including its longitudinal gauge non-invariant part, we must implement gauge invariance by imposing Gauss' law on the Hilbert space,
\be
  \bra{\vc{\A}}\partial_j \hat E_j(\z) \ket{\Psi,t} =
  -i\partial_j\frac{\delta}{\delta \A_j(\z)}\Psi[\vc{\A},t] =
  \rho(\z)\Psi[\vc{\A},t]\; ,
\ee
where $\rho(\z)$ is the charge distribution of the fermionic fields.

We will frequently use the transverse ($T$) and longitudinal ($L$) projectors, which in momentum space are
\be
  T_{ij}(\pb) = \delta_{ij} - \frac{p_ip_j}{|\pb|^2},\qquad
  L_{ij}(\pb) = \frac{p_ip_j}{|\pb|^2}\; ,
\ee
and obey $T+L=1$, $TL=0$. Transverse fields and derivatives are defined in a natural way by
\be
	f^T_i(\pb) := T_{ij}(\pb)f_j(\pb)\; ,
\ee
and similarly for longitudinal parts.

\subsection{Dressed states in the abelian theory}\label{dress-sect}
The Lagrangian fermions are not gauge invariant. In order to discuss physical, gauge invariant objects we dress the fermions with a function of the gauge field, which we write $\exp(W[\vc{\A}])$. We will be interested in states with two heavy fermions, where the state takes the form \cite{Zarembo:1998xq}
\be
	\e^{W[\ssvc{\A}]}\ q(\x_2)q^\dagger(\x_1)\ket{0}\; ,
\ee
for $q$ and $q^\dagger$ the heavy fermions and $\ket{0}$ the vacuum. In this section we will construct the most general functional $W[\vc{\A}]$ which makes the state invariant under the residual gauge transformations of Weyl gauge \footnote{The dressing typically has two parts. The `non-minimal' part of the dressing is chosen such that the heavy fermions are static in time, although in Weyl gauge this piece becomes trivial. In this paper we focus on the `minimal' part of the dressing which ensures gauge invariance. See \cite{Bagan:1999jf,Bagan:1999jk} for more details.},
\be\begin{split}
	\A_j(\x) &\to \A_j^\lambda(\x)\equiv \A_j(\x)+\p_j\lambda(\x)\; ,\\
	q(\x) &\to \e^{-ie\lambda(\x)}q(\x),\quad q^\dagger(\x) \to \e^{ie\lambda(\x)} q^\dagger(\x)\; .
\end{split}\ee
These imply that $W[\vc{\A}]$ must transform as
\be
	W[\vc{\A}^\lambda] = W[\vc{\A}]+ie\,\lambda(\x_2)-ie\,\lambda(\x_1)\; .
\ee
Equivalently, Gauss' law implies
\be\label{Gauss-qq}
	-\frac{\partial}{\partial x^i}\frac{\partial}{\partial \A_i(\x)}\ W[\vc{\A}] = ie\, \delta^3(\x-\x_2)-ie\, \delta^3(\x-\x_1)\; ,
\ee
which constrains only that piece of $W[\vc{\A}]$ which is linear in $\vc{\A}$ to have longitudinal part 
\be
	ie\, \frac{1}{\nabla^2}\partial_j \A_j(\x_2)-ie\, \frac{1}{\nabla^2}\partial_j \A_j(\x_1)\;.
\ee
Here, $\A_j$ may be replaced with its longitudinal part $\A^L_j$. This is called the `Coulomb dressing', first proposed by Dirac for modelling the gauge invariant static electron~\cite{Dirac:1955uv}. The general dressing which gives a gauge invariant charge-anticharge state is therefore
\be\label{general}
	\exp\bigg[ie\, \frac{1}{\nabla^2}\partial_j \A_j(\x_2)-ie\, \frac{1}{\nabla^2}\partial_j \A_j(\x_1) + J[\vc{\A}^T]\bigg]\; ,
\ee
for arbitrary $J$. We will investigate some of the choices of $J[\vc{\A}^T]$ below.

\subsection{The axial state}
What other principles shall we use to construct our dressings? Let us make an educated guess as to what a gauge invariant state might look like. Taking a bottom-up approach, we could think of beginning with the ground state in the vacuum sector, i.e. in the presence of no matter, $\rho(\z)=0$. This is the lowest, gauge invariant, eigenstate of our Hamiltonian (\ref{U1H}) which is easily found to be given by the Gaussian
\be\begin{split}
	\Psi_0[\vc{\A}^T]&:= \text{Det}^{1/4}{\sqrt{-\nabla^2}}\\
	&\times \exp \bigg[-\frac{1}{2}\int\!\frac{\ud^3 p}{(2\pi)^3}\ \A_i(-\pb) |\pb|\,T_{ij}(\pb)\, \A_j(\pb)\bigg]\; ,
\end{split}\ee
Gauss' law implying that this state is independent of $\A^L_j$ \footnote{With regard to the norm of the vacuum state, $\bracket{\Psi_0}{\Psi_0}$, the transverse integral is correctly normalised, the integral over $\ssvc{\A}^L$ contributes the infinite volume of the gauge group, which we should divide out. We therefore neglect such factors in the sequel.}. Now we add our fermion-antifermion pair, and ask what else we must add to make them gauge invariant.  A common approach is to connect the sources by a string. This gives us what we call the `axially dressed' state $\chi$,
\be\label{the-ax}
	\chi[\vc{\A}] = \exp\bigg[\, ie\int\limits_C\!\ud z^i \A_i(\z)\bigg]\Psi_0[\vc{\A}^T]\; .
\ee
We will take the curve $C$ to be the straight line from $\x_1$ to $\x_2$. Fourier transforming, it is easily checked that $\chi$ is of the general form (\ref{general}), with $J$ being the line integral of $\vc{\A}^T$. Therefore $\chi[\vc{\A}]\,q(\x_2)q^\dagger(\x_1)\ket{0}$ describes a gauge invariant state.

Have we found the ground state in the fermion-antifermion sector? Far from it -- the axially dressed state (\ref{the-ax}) is not even an eigenstate of the Hamiltonian, which, imposing a large momentum cutoff $|\pb|<\Lambda$,  has expectation value
\be\label{confy}
	\bra{\chi}\hat H\ket{\chi} =\epsilon_0 -\frac{e^2\Lambda}{4\pi^2} + \frac{e^2}{4\pi} \Lambda^2|\x_2-\x_1|+\ldots
\ee
where the ellipses denote terms vanishing as $\Lambda\to\infty$. The first terms denote the vacuum energy,
\be \label{eq:vacuumenergy}
	\epsilon_0:= \int\!\ud^3 x\,\int\!\frac{\ud^3 p}{(2\pi)^3}\ |\pb|\; ,
\ee
which obviously represents the total energy of a noninteracting photon `gas' filling all space,  and self-energies of the sources. The final term in (\ref{confy}) gives a linearly rising potential between the sources -- i.e. they appear to be confined! The cutoff $\Lambda^2$ regulates a short distance divergence $\delta^2(0)$ coming from the infinitely small extension of the string in the two directions perpendicular to $\n$. The potential diverges as we remove the cutoff, and the axially dressed state is therefore infinitely excited \cite{Haagensen:1997pi}.

In the classical theory it is known that the string between the sources radiates energy and decays in time to an energetically more favourable state describing sources surrounded by a Coulomb field \cite{Prokhorov:1992ry}. We will consider the quantum time development of this state below, but in preparation we first construct the true ground state.
\subsection{The ground state}
Gauss' law has already fixed the longitudinal piece of our wavefunctionals to be the Coulomb dressing, which is an eigenstate of the longitudinal part of the Hamiltonian. The lowest eigenvalue of the transverse part of the Hamiltonian is just the vacuum functional $\Psi_0[\vc{\A}]$. The ground state, or Coulomb dressed state, $\Phi$ in the fermion-antifermion sector is therefore \cite{Rossi:1979jf, Jackiw:1983nv, Zarembo:1997ms}
\be\label{TheGS}
	\Phi[\vc{\A}] \equiv\exp\bigg[\,ie\, \frac{1}{\nabla^2}\partial_j\A_j(\x_2)-ie\, \frac{1}{\nabla^2}\partial_j\A_j(\x_1)\bigg]\ \Psi_0[\vc{\A}^T]\; ,
\ee
with $\Phi[\vc{\A}]\, q(\x_2)q^\dagger(\x_1)$ describing a static, gauge invariant charge and anti-charge of minimal energy. The total energy of the ground state is found to be IR finite but UV divergent. For large values of the momentum cutoff we find 
\be\label{Ecoul}
	E_C :=\epsilon_0 + \frac{e^2\Lambda}{2\pi^2}-\frac{e^2}{4\pi|\x_2-\x_1|}\,\bigg(1+\mathcal{O}\big(\Lambda^{-1}\big)\bigg)\; ,
\ee
Again we have the vacuum and self energies. The final term is finite as the cutoff is removed and gives the Coulomb potential between the sources. The ground state in the fermion-antifermion sector is therefore described by two individually gauge invariant sources surrounded by a Coulomb field.

Our dressed states (\ref{the-ax}) and (\ref{TheGS}) are related by
\be
	\chi[\vc{\A}] =\exp\bigg[ ie\int\limits_C\!\ud z^i \A^T_i(\z)\bigg]\ \Phi[\vc{\A}]\; .
\ee
It is this extra transverse  piece which, in the classical theory, radiates away. We now study the time development of our quantum states. In our functional picture, time evolution described by the Schr\"odinger equation is implemented by taking the inner product of states with the Schr\"odinger functional. For the explicit form of this functional see, e.g. \cite{Rossi:1979jf,Luscher:1992an}. Here, the calculation reduces to performing a Gaussian integral, with the result that the axial state at time $t$ is
\bw
\begin{eqnarray}\label{evolved}
	\chi[\vc{\A},t] & =&\ \e^{-iE_Ct}\ \Phi[\vc{\A}]\ \exp\bigg[\frac{1}{2}\mathcal{N}(t)-\frac{1}{2}\mathcal{N}(0) +e\int\!\frac{\ud^3 p}{(2\pi)^3}\
\e^{-i|\pb|t}\bigg(\frac{\e^{i\pb{\cdot}\x_2}-\e^{i\pb{\cdot}\x_1}}{\n{\cdot}\pb}\bigg)\, n_k\A^T_k(\pb) \bigg]\; ,\\
	\mathcal{N}(t)&:=&e^2\int\!\frac{\ud^3 p}{(2\pi)^3}\ \frac{1}{|\pb|}\e^{-2i|\pb|t}\bigg(\frac{1-\cos{\pb{\cdot}(\x_2-\x_1)}}{(\n{\cdot}\pb)^2}\bigg)\, n_iT_{ij}(\pb)n_j\; ,
\end{eqnarray}
\ew
which defines $\mathcal{N}(t)$. There is no divergence at $\n{\cdot}\pb=0$ in any of our expressions, as the difference of exponentials, above, always appears to `regulate' the notorious axial gauge divergence \cite{Kalloniatis:1989qq}.

The axially dressed state, at time $t$, is comprised of the Coulomb dressed state together with an additional transverse contribution and a time dependent normalisation $\mathcal{N}(t)$. Imposing a momentum cutoff, the transverse term and $\mathcal{N}(t)$ are regulated expressions which are suppressed by rapid oscillations as $t\to\infty$. Alternatively, we
may use Abel's limit \cite{Weinberg:1995mt, Abel2},
\be
	\lim_{t\to\infty} f(t) = \lim_{\epsilon\to 0^+} \epsilon\int\limits_0^\infty\ud s\ f(s)\ \e^{-\epsilon s}\; ,
\ee
to study the large time behaviour of these oscillatory expressions. Physically, this limiting procedure adds a tiny imaginary part to the energies in the Fourier representation of $f$ which leads to exponential damping and the appearance of imaginary energy denominators. Applying this, for example, to the transverse term we take the limit
\be
	\lim_{\epsilon\to 0^+}\ \int\!\frac{\ud^3 p}{(2\pi)^3}\ \frac{-i\epsilon}{|\pb|-i\epsilon}\bigg(\frac{\e^{i\pb{\cdot}\x_2}-\e^{i\pb{\cdot}\x_1}}{\n{\cdot}\pb}\bigg)\, n_k\A^T_k(p)\; .
\ee
This is very similar to our original axial dressing but with improved UV convergence. As $\epsilon\to 0$ this expression vanishes, and the transverse contribution, along with $\mathcal{N}(t)$, drops out. Therefore, pre-multiplying by the exponential of the Coulombic energy, we find (with a chosen regularisation understood)
\be\label{limit}
	\lim_{t\to\infty}\ \e^{iE_Ct} \chi[\vc{\A},t] =\e^{-\mathcal{N}(0)/2}\ \Phi[\vc{\A}]\; .
\ee
This result echoes that of the classical theory. We have that both classically and quantum mechanically the axially dressed state decays to the Coulomb state in the large time limit. The additional factor $\exp(-\mathcal{N}(0)/2)$ ensures
conservation of probability.

\subsection{Confined and deconfined U(1) charges}
We have seen that the lowest energy state of a gauge invariant fermion--antifermion pair in U(1) gauge theory is given by the Coulomb dressing, the inter-fermion potential being the expected Coulomb potential. Connecting the fermions by a string gives an unphysical, infinitely excited state. It describes a confining potential between the matter
sources and decays to the energetically favourable ground state.

Having found the ground state exactly in this simple theory, we may ask what use, if any, has the infinitely excited axial state? To answer this, note that the Coulomb state $\Phi[\A]q(\x_2)q^\dagger(\x_1)\ket{0}$ describes two individually gauge invariant fermions. This may be seen from (\ref{TheGS}), as the Coulomb dressing factorises into two parts,
\bw
\be\label{2parts}
	\Phi[\vc{\A}] = \exp\bigg[\,ie\, \frac{1}{\nabla^2}\partial_j\A_j(\x_2)\bigg]\exp\bigg[-ie\, \frac{1}{\nabla^2}\partial_j\A_j(\x_1)\bigg]\Psi_0[\vc{\A}^T]\; .
\ee
\ew
The first exponential dresses $q(\x_2)$, giving us a gauge invariant fermion, and the second exponential makes $q^\dagger(\x_1)$ gauge invariant. In a deconfined theory this is the situation we expect --- individual charges exist and are separately gauge invariant.

It is not possible to unambiguously factorise the axial state into, similarly, one dressing for the charge and one for the anticharge. Taking (\ref{the-ax}), we Fourier transform and separate $\vc{\A}$ into transverse and longitudinal parts, writing the line integral as 
\be\begin{split}\label{factorno}
ie\, \int\!\frac{\ud^3p}{(2\pi)^3}\ &\bigg(\frac{\e^{i\pb{\cdot}\x_2}-\e^{i\pb{\cdot}\x_1}}{i |\pb|^2}\bigg)\,p_k\A_k^L(p) \\
&+ \bigg(\frac{\e^{i\pb{\cdot}\x_2}-\e^{i\pb{\cdot}\x_1}}{i \n{\cdot}\pb}\bigg)\,n_k\A_k^T(p)\; .
\end{split}\ee
We have written out the longitudinal projector explicitly. The first term is precisely the Coulomb dressing, as is necessary for gauge invariance. However, it is not possible to factorise the second, transverse, part of (\ref{factorno}) in a fashion similar to (\ref{2parts}) without introducing new divergences. For example, we could try to write,
\be\label{badsplit}
	\frac{\e^{i\pb{\cdot}\x_2}-\e^{i\pb{\cdot}\x_1}}{i \n{\cdot}\pb} \to \frac{\e^{i\pb{\cdot}\x_2}}{i\, \n{\cdot}\pb}-\frac{\e^{i\pb{\cdot}\x_1}}{i\,\n{\cdot}\pb }\; ,
\ee
but whereas the left hand side of this expression is regular everywhere the right hand side requires the regularisation of axial gauge poles.

What is the physical significance of this result? We saw earlier that the axial dressing appeared to lead to a confining potential. Physically, we therefore interpret the lack of a natural factorisation as an absence of individually gauge invariant electrons and positrons. Instead, there is only a single, neutrally charged, object bound by a (thin) flux tube. It would seem, then, that in a confining theory our axial state could be a better model of the true ground state than the Coulomb state.

A suitable toy model of such a confining theory is compact QED, originally introduced by Polyakov \cite{Polyakov:1975rs} who argued that the 4$d$ lattice model has a confinement-deconfinement phase transition. This was later proven analytically \cite{Banks:1977cc,Peskin:1977kp,Guth:1979gz},  see also \cite{Kogut:1982ds} and \cite{Seiler:1982pw}. In the confining phase we expect the electric field to be concentrated along a flux tube connecting the sources. Qualitatively, this does not look much like the Coulombic field which we have seen is optimal so far, so the ground state could well differ significantly from the simple functional (\ref{TheGS}).

Although in this phase neither of our states will represent the true ground state we may use them to extract information about it. We describe this method below and proceed to apply it to compact QED in the following section.

\subsection{A ground state probe}
What object are we to calculate which probes the ground state? Rotating to Euclidean space we consider, for any two states $\ket{\psi}$ and $\ket{\psi'}$, the ratio
\be
	\frac{\bra{\psi'}\e^{-\hat{H}t}\ket{\psi'}}{\bra{\psi}\e^{-\hat{H}t}\ket{\psi}}\; .
\ee
In the large time limit contributions from excited states are exponentially suppressed. This limit then projects onto the ground state, call it $\ket{0}$, so that our ratio tends to
\be
	\frac{|\bracket{\psi'}{0}|^2}{|\bracket{\psi}{0}|^2}\; .
\ee
If the ratio is less than one in the limit then $\psi$ has the better overlap with the ground state, whereas if the ratio is greater than one $\psi'$ is closer to the ground state. 

Below we will examine the ratio for $\psi'=\chi$ and  $\psi=\Phi$, our axial and Coulomb states respectively. Writing $r\equiv |x_2-x_1|$ the ratio of interest to us is
\be\label{theratio}
	R(r,t) := \frac{\bra{\chi}\e^{-\hat{H}t}\ket{\chi}}{\bra{\Phi}\e^{-\hat{H}t}\ket{\Phi}}\; ,
\ee
where the $r$ dependence is contained in the states. In the deconfined U(1) theory we have studied so far is
\bw
\be\begin{split}\label{explicit}
	R(r,t) &= \e^{E_Ct}\bra{\chi}\e^{-\hat{H}t}\ket{\chi}\; , \\
	&=\exp\bigg[-e^2\int\limits^\Lambda\!\frac{\ud^3 p}{(2\pi)^3}\
\frac{1}{|\pb|}(1-\e^{-|\pb|t})\bigg(\frac{1-\cos (r\, \n{\cdot}\pb)}{(\n{\cdot}\pb)^2}\bigg)n_iT_{ij}(\pb)n_j\bigg]\; , \\
	&\to \exp(-\mathcal{N}(0))\quad\text{as $t\to\infty$}\; ,
\end{split}\ee
\ew
where the first line follows from $\Phi$ being an eigenstate.

In the large time limit this correctly reproduces the mod squared overlap between the axial state and the ground (Coulomb) state,
\be\label{N}
	\bracket{\chi}{\Phi} = \exp\big( -\mathcal{N}(0)/2\big)\;.
\ee
$\mathcal{N}(0)$ is UV divergent (so that the overlap formally vanishes, as one may expect for the overlap between a physical and an unphysical state), but with the UV regulator in place $\mathcal{N}(0)$ is positive definite and so our ratio becomes $|\bracket{\chi}{\Phi}|^2<1$, as we would expect since $\ket{\Phi}$ is the ground state and $\ket{\chi}$ is not.

\begin{figure}
\includegraphics[width=7cm]{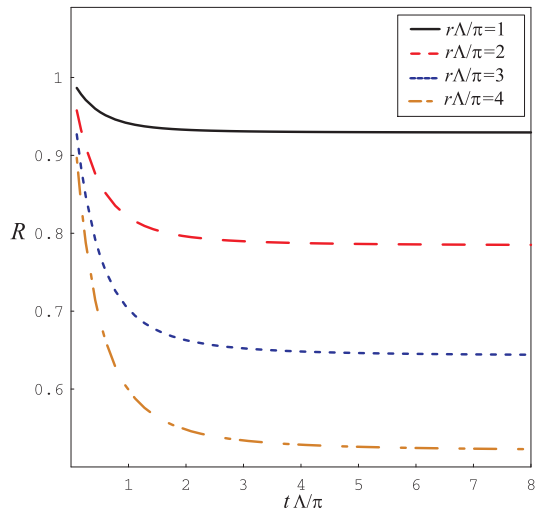}
\caption{The ratio $R(r,t)$ defined in (\ref{theratio}) for unconfined U(1). Units are discussed below (\ref{explicit}). As the separation of the charges increases the axial state becomes an increasingly poorer description of the ground state.}
\label{fig:Analytic1}
\end{figure}

The ratio is plotted in Figure \ref{fig:Analytic1}. To make contact with the lattice data of the following section, we take $e^{-2}=1.05$ and measure distance in units of $\pi/\Lambda$. In these dimensionless units momentum is cut off at $|\pb|<\pi$ so that $1/\Lambda$ is associated with the lattice spacing. Increasing the value of the cutoff does not qualitatively change the features of the plots. The integrand above is positive semi definite so that (with the regulator in place), we have $R(r,t)<1$ for all $r$ and $t$, as it must be since $\Phi$ is the ground state and $\chi$ is not. This may be seen in Figure \ref{fig:Analytic1}, which also shows that the ratio decreases rapidly with $t$ to its asymptotic value of $\exp(-\mathcal{N}(0))$. We also observe that the ratio decreases as we increase the separation $r$ of the charges. Increasing the length of the string concentrates more and more energy along the lengthening flux tube, giving an increasingly worse description of the true $1/r$ falloff of the Coulomb field.

We turn now to the confining phase.

\section{A simple confining model: compact U(1)}\label{U1sect}
There is no exact analytic treatment of compact QED, although we note that the functional methods we employ have been successfully applied in $2+1$ dimensions \cite{Kogan:1994vb, Nolland:2004cv}. We therefore turn to lattice methods. In the following we outline our numerical approach and go on to calculate the ratio $R(r,t)$. In the deconfined phase we will see that our analytic results can be reproduced numerically, providing a check on our methods. We will then go on to study the confining phase.
\subsection{Setting up compact QED on the lattice}
In the lattice formulation, the degrees of freedom are the fields
$$
U_{\mu}(x) = \exp(i \theta_{\mu}(x))\; , \qquad -\pi< \theta _\mu (x) \leq \pi\;,
$$
which are associated with
the links of the  4--dimensional Euclidean space-time lattice.
Using the standard Wilson action, the partition function is given by
\be
Z \; = \; \int {\cal D} \theta _\mu \; \exp \Bigl\{
\beta\, \sum_x\sum_{\mu < \nu} \cos \left(\theta_{P}(x)_{\mu \nu}\right)  \Bigr\} \; ,
\ee
where $\beta=1/e^2$ with $e$ the bare gauge coupling.
The plaquette angles are defined by
\be
    \theta_{P}(x)_{\mu \nu} = \theta_{\mu}(x) + \theta_{\nu}(x
+ \mu) - \theta_{\mu}(x + \nu)   -  \theta_{\nu}(x) \; .
\ee
Because of the compact domain of support for the degrees of freedom,
the U(1) theory admits magnetic monopoles. High precision measurements indicate that at $\beta=\beta_\text{crit}$, with $\beta_\text{crit}$ = 1.0111331(15)
in $3+1$ dimensions~\cite{Vettorazzo:2004cr}, there is a phase transition. Below the transition, $\beta<\beta_\text{crit}$, there is an abundance of magnetic monopoles leading  to confinement of electric charges by means of the dual Meissner  effect \cite{Polyakov:1975rs}. This has been convincingly confirmed by lattice simulations \cite{DeGrand:1980eq,Stack:1991zp}. For $\beta \ge \beta_\text{crit}$, monopole nucleation is suppressed, and the theory is realised in the standard Coulomb phase.

Our lattice comprises 12$^{4}$ points at $\beta$=1.0 (confining phase) and $\beta$=1.05 (Coulomb phase). A subtlety is that we must use open boundary conditions in the spatial directions and periodic boundary conditions in the temporal direction. This particular setup is necessary for a proper implementation of the axial gauge, further discussed below. In order to reduce the influence of boundary effects, we always allow a distance of two lattice spacings to the boundaries when measuring observables.

A standard heat-bath algorithm combined with micro-canonical reflections
was used to bring the initial configurations into thermal equilibrium.
Both disordered (hot) and ordered (cold) initial configurations were
used and it was verified that our measurements were independent
of the choice of the initial configuration. Measurements
were taken with 20,000 configurations for each of the two phases.

\subsection{Gauge fixing: Coulomb and axial gauges}
The quantities of interest are correlators of short Polyakov lines of temporal extent $t$ and spatial separation $r$, which begin in the time slice $x^0=0$ and end in the time slice $x^0=t$. Bringing these correlators into specific gauges amounts to choosing specific dressings for the static test charge (and corresponding anti-charge). For more details see \cite{Heinzl:2007cp}.

Let us focus first on Coulomb gauge fixing. Given the gauge
transformation $\Omega(x)=\exp(i\alpha(x))$,
\be\begin{split}
	U_\mu (x) \to U^\Omega _\mu (x) &=  \exp \big( - i\alpha (x+\mu)+i\alpha (x) \big)\
U_\mu (x) \; ,\\
 \quad &-\pi<\alpha (x)\leq\pi\; ,
\end{split}\ee
Coulomb gauge fixing is implemented by maximising, with respect to $\alpha$, the gauge fixing functional
\be
    \mathcal{F}[U^\Omega] = \frac{1}{2}\ \text{Re} \bigg[ \sum_{x}
\sum^{3}_{i=1} U^\Omega_{i}(x) \bigg]\; .
\ee
We use a standard iteration-overrelaxation scheme for this task.  To provide a stopping criterion for the iteration, we define
\be\begin{split}
\Delta &= \frac{1}{N_\mathrm{in}} \sum_{x_\mathrm{in}} \Delta^{2}_{x}\;, \\
\Delta_{x} &= \text{Im} \sum^{3}_{i=1}\big[
U^\Omega_{i}(x) + U^{\Omega \, \dagger}_{i}(x-i) \big]\; ,
\end{split}\ee
where the sum over (interior) lattice points excludes points on the boundary, and where consequently $ N_\mathrm{in} = (N_{i}-2)^{3}\times N_{t} $.
Introducing the photon field $A_\mu (x)$ via
\be
	U_\mu(x) \; = \; \exp \big( i a \, A_\mu(x) \big) \; ,
\ee
the above quantity $\Delta $ is a measure of the violation of the
transversality condition, as can be seen from
\be
	\Delta = \frac{a^4}{N_\text{in}} \sum_{x} \big[ \partial_{i} A_{i}(x)\big]^{2} +
\mathcal{O}(a^5) \; .
\ee
We use $ \Delta  <  10^{-12}$ in our simulations.

While Coulomb gauge fixing involves a non-linear optimisation problem,
axial gauge fixing can be implemented straightforwardly.
The axial gauge condition is given by
\be
	A_3(x)= 0 \; , \qquad\text{equivalently}\  U_3^\Omega (x)=1 \; ,
\ee
choosing our earlier unit vector $\n$ to point in the 3--direction. Let us assume that we have $N_3$ lattice points in the 3--direction, and the gauge transformations are numbered according to
\be
	\Omega _k \equiv \Omega (x_k)\, \quad\text{with}\ x_k\equiv (x^1,x^2,ka,x^4)\; .
\ee
The crucial observation is that, because of open boundary conditions,
the gauge transformations $\Omega _1$ and $\Omega _{N_3}$ are
independent. Hence, it is always possible to choose $\Omega _{k+1}$
to satisfy
\be
	U_3^\Omega (x_k) =\Omega_k \, U_3 (x_k) \, \Omega^\dagger_{k+1} =1
\; .
\ee
As a result $\Omega _1$ remains undetermined and represents a residual
gauge degree of freedom.

\subsection{Probing the ground state of compact QED}

Now that we have our lattice implementation of the dressings we may investigate their physical significance by looking at the overlap of the dressed states with the ground state.
We will calculate the ratio $R(r,t)$, defined in (\ref{theratio}), which probes the ground state in the large $t$ limit. It is useful to introduce the matrix of transition amplitudes
\be
    M(r,t)=
    \begin{pmatrix}
        C_{\Phi\Phi}(r,t)& C_{\chi\Phi}(r,t)\\
        C_{\Phi\chi}(r,t)& C_{\chi\chi}(r,t)
\end{pmatrix}\; ,
\ee
where $C_{\psi'\psi}:=\bra{\psi'}\e^{-\hat{H}t}\ket{\psi}$. This enables us to determine (for fixed values of $r$) the minimal value of $t$ for which the contributions of excited states to the $C_{\psi'\psi}(r,t)$ are sufficiently suppressed so that we can extract the overlaps with the ground state. Inserting a complete set of states, we obtain
\be
    C_{\psi'\psi} = \sum_{n} \e^{-E_{n}t} \bracket{\psi'}{n}\bracket{n}{\psi}\; ,
\ee
where as usual $E_{n}$ denotes the eigenvalue of $\hat{H}$ for the state $\ket{n}$. All contributions but that of the ground state ($n=0$) to $C_{\psi'\psi}$ can be neglected due to exponential suppression in the large $t$ limit. A calculation of the determinant of the
matrix $M$ yields the large time limit
\be
e^{2E_0t}\det[M(r,t)] \to \left| \begin{array}{cc}
        |\bracket{\Phi}{0}|^2& \bracket{\chi}{0}\bracket{0}{\Phi} \\
        \bracket{\Phi}{0}\bracket{0}{\chi} & |\bracket{\chi}{0}|^2
\end{array}\right| =0\; .
\ee
Therefore, if the value of the determinant of $M$ substantially deviates from zero (within error bars), the large $t$ limit has not yet been reached.

The calculation of $C_{\Phi\Phi}(r,t)$, the Coulomb
dressed Polyakov line correlator (pictorially, we may represent this object by $|\ \, |$, see \cite{Heinzl:2007cp}), is straightforward.
Calculating its axially dressed counterpart amounts to
calculating the ordinary, gauge invariant, Wilson loop. This can be
seen as follows: our starting point is the correlator of two
Polyakov lines $P(\x_1,\x_2)$ of temporal extent $t$ and distance $r$
calculated from gauge fixed links
\be
P(\x_1,\x_2)[U^\Omega _\mu] \; , \hbo \x_2 = \x_1 + r \n \; ,
\ee
where $\n$ is the unit vector in 3--direction.
Since in axial gauge the links $U_3^\Omega $ are gauged to the unit
elements, the above correlator can be completed to an $r \times t$
Wilson loop lying in the $3{-}t$ plane. Since the Wilson loop is manifestly
gauge invariant, we find:
\be
P(\x_1,\x_2)[U^\Omega _\mu] = W[U^\Omega _\mu] = W[U_\mu ] \; .
\ee
In other words, the loose ends of the
Polyakov lines are joined by a straight line of
parallel transporters $U_{\mu}(x)$ in the initial and final time slices
(this may be represented by $\Box$). Calculating the correlator
of charges that evolve from axial gauge into Coulomb gauge,
$C_{\Phi\chi}(r,t)$, means putting in a straight line of
parallel transporters between the static charges in the initial time
slice, the resulting object being of a $\sqcup$ shape. Calculating the
correlator from Coulomb to axially dressed charges $C_{\chi\Phi}(r,t)$
proceeds similarly, looking like a staple, $\sqcap$.

\subsection{Numerical results}
\begin{figure}
\includegraphics[width=7cm]{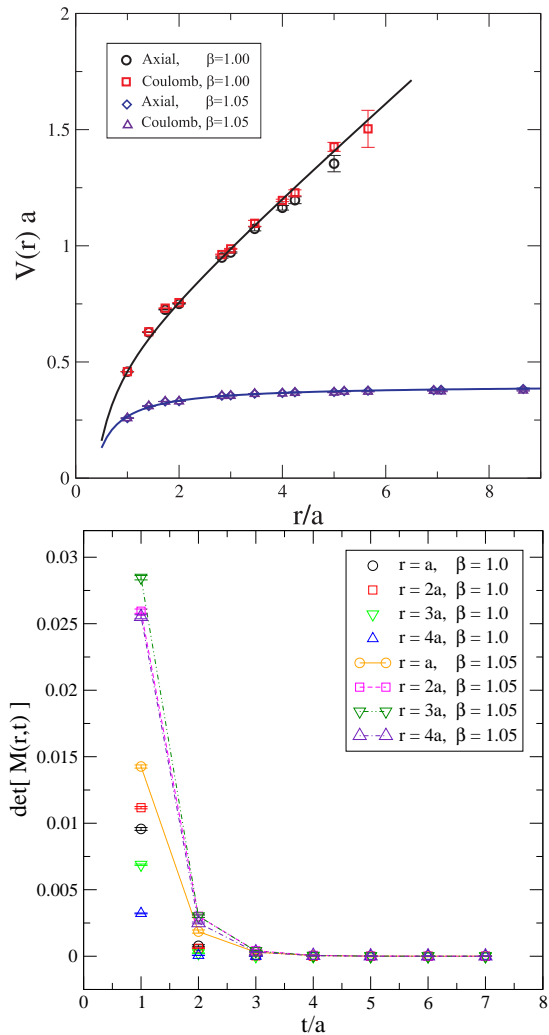}\hspace{0.5cm}\includegraphics[width=7.2cm]{F3-Determinant.eps}
\caption{Upper panel: the U(1) inter-fermion potential in the confined and deconfined phases. Lower panel: behaviour of the determinant $\text{det}[M(r,t)]$.}
\label{fig:potent-u1}
\end{figure}

Calculating the familiar static potential between both axially and Coulomb dressed charges below and above $\beta_\text{crit}$, the phase transition is clearly illustrated. These results are displayed in Figure \ref{fig:potent-u1} (upper panel). We see the Coulomb potential above the transition, $\beta=1.05$, and the linear rising potential below the phase transition, $\beta=1.0$ (see also \cite{Panero:2005iu} for numerical investigations of the confining phase).
In the lower panel of Figure \ref{fig:potent-u1} we have plotted the behaviour of the determinant $\text{det}[M(r,t)]$. We see that the determinant vanishes rapidly, coming effectively to zero at around $t=3a$ both above and below the phase transition. Thus the `large time limit', in which we can extract good overlaps with the ground state, is reached quite rapidly.

\begin{figure}
\includegraphics[width=7cm]{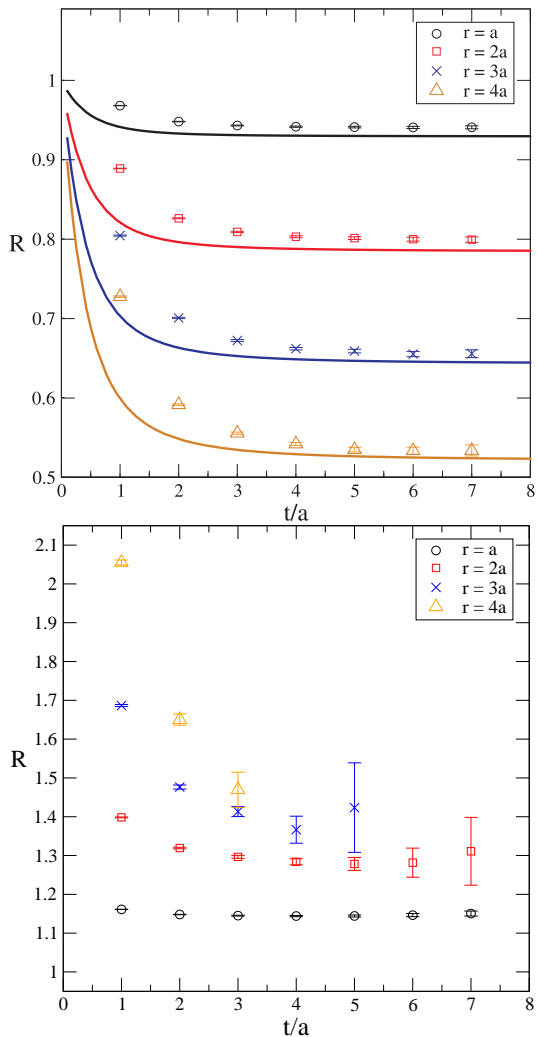}\hspace{0.5cm}\includegraphics[width=7cm]{F5-CorrRatios.eps}
\caption{The overlap ratio $R(r,t)$ as a function of $t$ and $r$ in the deconfined/Coulomb phase, above the phase transition (upper panel, solid lines are analytic results) and in the confining phase, below the phase transition (lower panel). Lattice: 12$^{4}$, spatially open boundary conditions, temporally periodic boundary conditions, $\beta $ = 1.05 (upper panel) and $\beta $ = 1.0 (lower panel). }
\label{fig:wolfu1}
\end{figure}

The numerical results for the overlap ratio $R(r,t)$ are shown in Figure \ref{fig:wolfu1}. The upper panel shows the Coulomb phase results ($\beta$ =1.05). Figure \ref{fig:Analytic1} is overlaid for comparison. We see a close agreement between the analytic solution and the lattice data, despite the different short distance cutoffs in use (the lattice spacing $a$ in each direction compared to the $O(3)$ symmetric momentum cutoff $|\pb|<\Lambda$). We observe the formation of a plateau at around $t=3a$, where the ratio approaches its asymptotic value, in agreement with the behaviour of the determinant of $M(r,t)$, cf.~Figure \ref{fig:potent-u1}. Increasing the spatial separation $r$ of the charges probes the physically more interesting long-distance behaviour. In the Coulomb phase, as seen in the analytic calculations, the ratio $R(r,t)$ decreases as the separation $r$ of the charges is increased. The energy of the axial state increases linearly with $r$, giving an increasingly worse overlap with the ground (Coulomb) state.

Encouraged by the agreement between our numerical and analytic results, we now consider the confining phase, $\beta=1.0$. The results are shown in the lower panel of Figure \ref{fig:wolfu1}. Most notably, the ratio is now greater than one, which signals that the axially dressed state has the better overlap with the ground state. Although the signal is rather noisy in this phase, compared to the Coulomb phase, one can still observe the formation of a plateau in $R(r,t)$ up to $r=3a$. Beyond this value, the statistics were not sufficient to extract data points which met our acceptance criterion of a relative error smaller than 10\%. As $r$ increases the ratio $R$ rises well above 1. There is no string breaking, and for larger $r$ the Coulomb potential, falling off as $1/r$, becomes a poorer and poorer description of the long flux tube connecting the sources.
These results show that the ground states differ significantly above and below the phase transition. Our physical intuition tells us that in a deconfined theory we expect to find two separate, gauge invariant charges. In a confining theory, however, we presume that physical states should be composite, uncharged objects. These expectations are indeed supported by our results. In the confining phase the Coulomb state, with its individual physical charges,  is not a good description of the ground state. Instead it is now the, overall uncharged, axial state with its narrow flux tube (its divergences regulated by the lattice spacing) which more closely resembles the true physics.

In the next section we will extend our discussion to non-abelian theories.

\section{Towards QCD: the ground state in the SU(2) Higgs model}\label{SU2sect}
Ideally, we would like to study the overlaps of the
Coulomb and axially dressed states with the true
ground state in a non-abelian pure gauge theory (no matter) which has both a confining and a non-confining phase. However, for an SU(N) pure gauge theory asymptotic freedom implies that there
is no fixed point in coupling constant space other than that at
$\beta \to \infty $. Hence, the SU(N) pure gauge theory has
only a single, confining, phase at zero temperature. High
temperature SU(N) Yang-Mills theory should provide a
non-confining framework. However, as the critical
temperature of, e.g., SU(2) pure gauge theory is
approximately $T_c \approx 300 \,$MeV, the
temporal extent of the lattice is
\be
	L_t \; \approx \; \frac{1}{300 \, \hbox{MeV} } \; \approx \;
\frac{2}{3} \, \mathrm{fm}\; .
\ee
Hence, excited states are only suppressed by a factor $\exp ( - H \, L_t )$ which is not enough to provide a good overlap with only the ground state.

The simplest theory which provides both a confining and a
non-confining phase at zero temperature is an SU(2) gauge
theory with a Higgs field $\phi $ in the fundamental
representation. This will be our model, but before we can study it we must briefly discuss what we mean by the Coulomb and axial dressings in a non-abelian theory.

\subsection{Non-abelian dressings}
As in Section \ref{analytic}, we are interested in the properties of quark-antiquark states made gauge invariant by using dressings. The form of the non-abelian dressings is much more complex than their abelian counterparts. However, in a perturbative expansion the lowest order term of the non-abelian dressings is typically given by taking the dressings of Section \ref{analytic} and replacing $\vc{\A}_j$ with $\vc{\A}^a_j\mathbf{T}^a$ for $\mathbf{T}^a$ the Lie algebra generators. For example, the Coulomb dressing is, to lowest order in the coupling $g$,
\be
	\exp\bigg(g\frac{\partial_j\A_j^a \mathbf{T}^a}{\nabla^2}+\ldots\bigg)\; ,
\ee
of similar form to the abelian case, (\ref{general}). All orders in this dressing are required to make gauge invariant states, and for the construction of higher order terms, and issues relating to Gribov copies, we refer the reader to \cite{Heinzl:2007cp, Lavelle:1995ty, Ilderton:2007qy}.

Just as in the abelian case, correlators between dressed states are calculated (using the full dressing) by putting time slices into particular gauges -- Coulomb and axial in our case. The methods employed in the U(1) theory may then equally be applied to non-abelian theories. With this understanding, we proceed to define and examine our SU(2) model.
\subsection{The SU(2) Higgs model}
We will study a gauged $O(2)$ model where the
length of the Higgs field is restricted to unity, \be \phi ^\dagger
\phi \; = \; 1 \; . \label{eq:on1} \ee It has long been known
that the confinement phase and the Higgs phase are smoothly
connected~\cite{Fradkin:1978dv}. A true order parameter, i.e., a
local operator which distinguishes between both phases, does
not exist. When static quark and antiquark sources are exposed
to the gluon field the electric flux tube between them may
break if the energy stored in the tube exceeds twice the
mass of the Higgs particle, and the string tension vanishes. Although a local order parameter does
not exist, the physics in the string breaking phase is
vastly different from that in the Higgs phase: while short
electric flux tubes form in the confining phase (thereby giving
rise to a ``string tension'' at intermediate distances), there
is no flux tube at all in the Higgs phase. Although this
behaviour cannot be detected with local order parameters,
non-local quantities, such as the vortex percolation
probability~\cite{Langfeld:2003zi}, can map out the phase
diagram. In fact the critical line for vortex percolation
follows the first order line of the phase diagram, and turns
into a Kert\'esz line in the crossover
regime~\cite{Langfeld:2002ic,Wenzel:2005nd,Langfeld:2004vu}.

The Higgs doublet
\be
	\phi = \left(\begin{array}{c} \phi _1 \\ \phi _2 \end{array} \right) \; ,\qquad\phi _{1,2} \in \mathbb{C}\; ,
\label{eq:on2}
\ee
may be equivalently written as a matrix $\alpha$,
\be
	\alpha = \left( \begin{array}{cc} \phi_1 &  - \phi^\ast _2 \\ \phi _2 & \phi^\ast _1 \end{array}\right) \; ,
\label{eq:on3}
\ee
where $\alpha$ is unitary. The lattice action of the theory is then
\begin{align}
	S &=S_\mathrm{Wilson} \; + \; S_\mathrm{Higgs} \; , \label{eq:on4} \\
	S_\mathrm{Wilson} &=\beta\! \sum _{x, \mu<\nu } \frac{1}{2} \;
\tr \; U_\mu (x) \, U_\nu (x+\mu) \, U^\dagger_\mu (x+\nu ) \,
U^\dagger _\nu (x) \; , \label{eq:on5} \\
	S_\mathrm{Higgs} &=\kappa \sum _{x, \mu }\ \frac{1}{2}\ \tr \; \alpha (x)\ U_\mu (x)\ \alpha ^\dagger (x + \mu ) \; ,
\label{eq:on6}
\end{align}
where $\kappa$ is the Higgs hopping parameter. The ``gluonic'' degrees of freedom are represented by the SU(2) link matrices $U_\mu (x)$ as usual. The action and partition function $Z$,
\be
	Z = \int{\cal D} U_\mu \; {\cal D} \alpha \; \exp\, (S) \; ,
\label{eq:on7}
\ee
are invariant under SU(2) gauge transformations:
\begin{align}
	U^\Omega _\mu (x) &= \Omega (x)\ U_\mu(x)\ \Omega^\dagger(x+\mu) \; , \qquad \Omega \in SU(2) \label{eq:on8} \\
	\alpha ^\Omega(x) &= \alpha (x) \; \Omega ^\dagger (x) \; .
\label{eq:on9}
\end{align}
In addition the theory possesses a global $O(2)$ symmetry:
	\be \alpha ^\omega (x) = \omega\, \alpha (x) \; ,\qquad \omega \in O(2) \; .
\label{eq:on10}
\ee
\begin{figure}
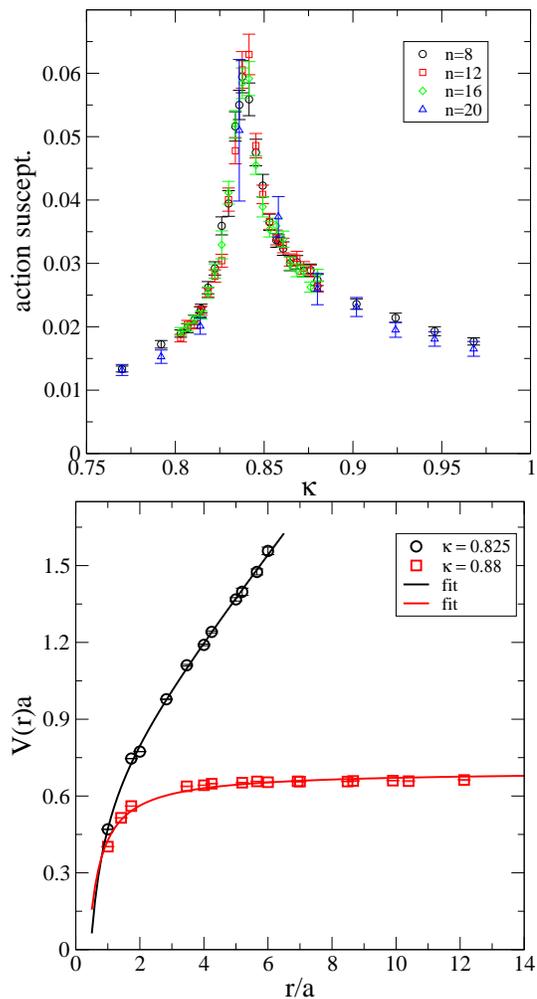

\includegraphics[width=7cm]{F6-suscept.eps}\hspace{0.5cm}\includegraphics[width=7cm]{F7-potential.eps}
\caption{
Upper panel: The susceptibility of the Higgs action (\ref{eq:on11}) as a function of $\kappa $ for $\beta =2.2$ for several lattice sizes $N$. Lower panel: The static quark potential for a $16^4$ lattice and $\beta =2.2$ for two different $\kappa $ values.}
\label{fig:on1}
\end{figure}
The model may be studied using standard lattice techniques. We work with $\beta=2.2$, several values of $\kappa$ and values of the lattice size $N$ \footnote{We do not work at a smaller value of $\beta$ in order to avoid lattice artifacts which dominate in the strong coupling, non universal, regime.}. The susceptibility of the Higgs action,
\be
	c_\mathrm{Higgs} = \frac{1}{N^4}\, \left[ \left\langle S_\mathrm{Higgs} ^2 \right\rangle - \left\langle S_\mathrm{Higgs} \right\rangle ^2\right] \; ,
\label{eq:on11}
\ee
may be used to explore the phase structure (for fixed $\beta $). Our numerical results are shown in the upper panel of
Figure~\ref{fig:on1}. The susceptibility is rather independent
of the lattice size and strongly peaked at a critical value $\kappa=\kappa_c$, where $\kappa _c  \approx
0.839(2)$. This indicates that the model undergoes a
crossover from the confinement phase to the Higgs phase at
$\kappa _c$.

Figure \ref{fig:on1}, lower panel, shows the static quark potential for $\kappa = 0.825$ (confined phase) and for $\kappa =0.88$ (Higgs phase)
which is significantly bigger than the critical value $\kappa_c$. In the
low $\kappa$ phase, we observe a linear rise of the potential
at large quark-antiquark distances. String breaking, induced
by the presence of the fundamental Higgs field, is not observed
for $r \, a \le 8$. We do, however, observe a string tension
\be
	\sigma\, a^2 \approx 0.15(1) \; ,\qquad\beta = 2.2 \; , \quad\kappa = 0.825 \; ,
\label{eq:on12}
\ee
that is somewhat reduced compared to the string tension of pure SU(2)
Yang-Mills theory, $\sigma a^2 \approx 0.28(1)$ (for $\beta =2.2$,
$\kappa =0$). In contrast, the potential data at $\kappa =0.88$
are well fitted by a Coulomb law.
\subsection{Overlaps with the ground state}
\begin{figure}
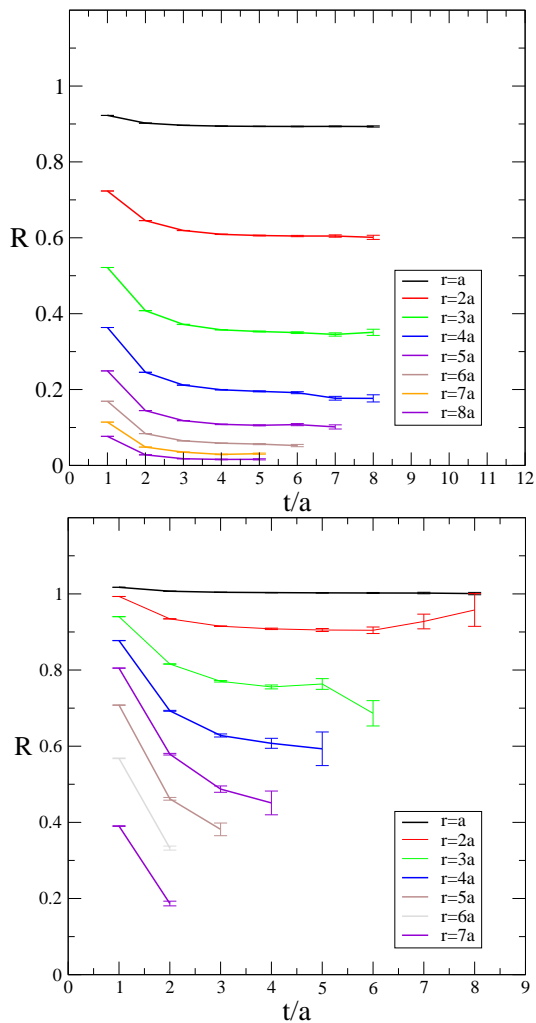

\includegraphics[width=7cm]{F8-overlap_80.eps}\hspace{0.6cm}\includegraphics[width=6.9cm]{F9-overlap_75.eps}
\caption{Ratio of overlaps between the axially dressed and the Coulomb dressed quark states in the Higgs phase ($\kappa=0.88$, upper panel) and the confining phase ($\kappa=0.825$, lower panel). Lines have been added between data points to guide the eye.}
\label{fig:on2}
\end{figure}

The ratio $R(r,t)$ in this theory is shown in Figure~\ref{fig:on2}. In the Higgs phase ($\kappa = 0.88$) the ratio is below one showing that
the Coulomb state has a better overlap than the axial state, similar to our U(1) results. Comparing the two panels of Figure~\ref{fig:on2} one sees that, for a given separation $r$, the ratio is slightly larger in the confined phase  ($\kappa = 0.825$) but it is still below one. This implies that, counter intuitively, the Coulomb dressed quarks still provide a better approximation to the true ground state.
We attribute this finding to a bad overlap between the axial and ground states caused by the infinitely thin flux tube of the axial state. The true ground state is expected to sustain a flux tube with a thickness of the order of several lattice spacings. In this regime, the ansatz of a very thin flux tube, whose potential energy is only finite because of the lattice spacing, is not as good a description as the Coulomb dressing.

\begin{figure}
\includegraphics[width=7cm]{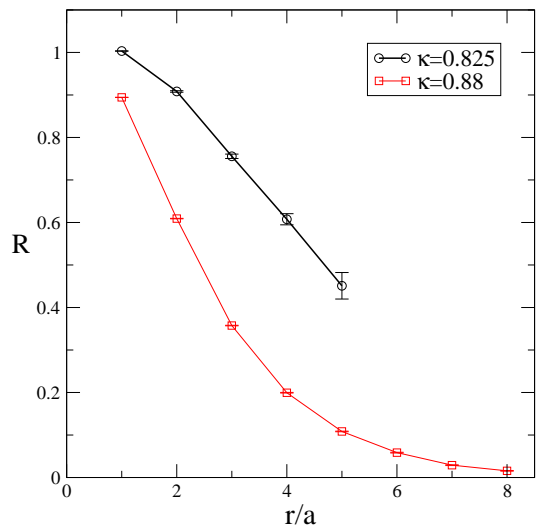}
\caption{Ratio of overlaps at $t=4a$, plotted as a function of the quark separation $r/a$ in the Higgs phase ($\kappa=0.88$) and the confining phase ($\kappa=0.825$).}
\label{fig:overlap_r}
\end{figure}

Our analysis suggests, then, that the traditional view of a thick flux tube \cite{Luscher:1980ac} is a better description of the true ground state.  In Figure \ref{fig:on2} we see that as the separation of the charges increases, the ratio $R$ decreases (at all times plotted), indicating that a thin string is a poorer description of the ground state at large $r$. To emphasise this point we plot, in Figure \ref{fig:overlap_r}, our ratio in the `large time limit', $t=4a$, where the ratio is clearly seen to decrease with increasing separation in both phases. This indicates, in particular, that in the confining phase, at fixed lattice spacing, the thin string is a poor description of the ground state at large separations.

\section{Conclusions}\label{concs}
In this paper we have studied various locally gauge invariant
ans\"atze for the lowest energy state in the heavy
fermion-antifermion sector of both abelian and non-abelian gauge
theories. One of these constructions corresponds to a Coulombic
dressing around each matter field, i.e., the state is made
up of two separate physical charges (with opposite signs). In the other, axial, ansatz the fermions are linked by a string like flux tube. This dressing does not factorise naturally, so no locally invariant charges can be defined. There is instead a single, uncharged object. We have compared how these two constructions overlap with the ground state for different theories and in various phases. Our primary tool here is the ratio $R(r,t)$, defined in (\ref{theratio}), which compares the overlaps between two ans\"atze and the true ground state.

Firstly we considered U(1) theory where we have a good understanding
and expect the ground state to correspond to two Coulomb charges.
Using the functional Schr\"odinger representation we were able to construct the
Coulomb state and also to show that the axial state is unstable and
decays into the Coulomb one. This was followed by lattice
simulations of compact U(1). In the deconfined phase the results
agreed very well with our analytic arguments, the Coulomb dressed
state being clearly preferred.

In the confined phase of compact U(1), where we do not have a good analytic understanding, our simulations showed that the axial state
was preferred. This is perhaps what one might expect, the axial
state corresponds to a flux tube between the fermions. However, it
is an extremely thin flux tube --- which is only non-zero on the
links between the fermions --- and in previous work \cite{Heinzl:2007cp} it was shown that for SU(2) a Coulombic dressed state more readily
yielded the interquark potential than an axial state. This
seems to indicate a major difference between the confining flux
tubes in compact U(1) and SU(2).

Spurred on by this we then studied the SU(2) Higgs model where there
is a crossover from a Higgs phase to a confined phase. We calculated
the potential on either side of the crossover and clearly saw the
two phases and the effect of the Higgs field on the potential
compared to the pure SU(2) theory. As with pure SU(2), \cite{Heinzl:2007cp}, our
simulations clearly demonstrated that the Coulomb state was
preferred to the axial ansatz on both sides of the crossover. We
interpret this as a consequence of the extreme thinness of the axial
flux tube. Presumably the flux tube in the true ground state in this
non-abelian theory is much thicker than its compact U(1) counterpart
and this is why the Coulomb dressing is preferred despite its slow
($1/r$) fall-off. It is interesting to speculate on the
phenomenological implications of this insight.

Our results agree with those of the investigations \cite{Boyko:2007ae, Boyko:2007jx}, where thickening of the tube with an increase in charge separation was observed at a fixed lattice spacing. However, those authors, based on an ansatz for the form of the string, also claimed that for finer lattices the width of the flux tube was proportional to the lattice spacing. This surprising result was argued to imply that in the continuum limit the width of the flux tube vanishes, i.e., the continuum ground state would be described by an infinitely thin string of flux. This result is the subject of some current debate and we will return to it in future work. 

The differences between the overlaps in the confining phases of the abelian and non-abelian theories clearly deserve further study. It would be interesting to try to construct different dressings which correspond to a thicker flux tube (\lq cigar shape\rq) around the fermions but which still fall off more rapidly than the Coulombic dressing. One possible way to approach this would be to use a dressing which corresponds to some form of interpolating gauge, somewhat analogous to ideas of `t~Hooft \cite{tHooft:1971rn,tHooft:1981ht} and others \cite{Chan:1985kf,Lavelle:1988fj,Heinzl:1990,Landshoff:1993da} on gauges depending on some `flow' parameter. A gauge interpolating between axial and Coulomb gauge, say, is expected to yield, for an optimal choice of the interpolating parameter, a still better overlap with the ground state. It could then be used for more efficient lattice calculations of, e.g., the interquark potential.

It would also be interesting to compare the overlap of axial and
Coulombic dressings in the SU(3) theory; this would let us
investigate whether modifying the number of colours significantly
alters the flux tube between two heavy quarks.

\acknowledgments
The numerical calculations in this paper were carried out on the HPC and PlymGrid facilities at the University of Plymouth.


\end{document}